# ELECTRON CLOUD EXPERIMENTS AT FERMILAB: FORMATION AND MITIGATION

R. Zwaska, Fermilab, Batavia, IL 60510, USA

*Abstract*

We have performed a series of experiments at Fermilab to explore the electron cloud phenomenon. The Main Injector will have its beam intensity increased four-fold in the Project X upgrade, and would be subject to instabilities from the electron cloud. We present measurements of the cloud formation in the Main Injector and experiments with materials for the mitigation of the Cloud. An experimental installation of Titanium-Nitride (TiN) coated beam pipes has been under study in the Main Injector since 2009; this material was directly compared to an adjacent stainless chamber through electron cloud measurement with Retarding Field Analyzers (RFAs). Over the long period of running we were able to observe the secondary electron yield (SEY) change and correlate it with electron fluence, establishing a conditioning history. Additionally, the installation has allowed measurement of the electron energy spectrum, comparison of instrumentation techniques, and energy-dependent behavior of the electron cloud. Finally, a new installation, developed in conjunction with Cornell and SLAC, will allow direct SEY measurement of material samples irradiated in the accelerator.

## INTRODUCTION

The electron cloud is a phenomenon that can affect any high current accelerator. It is characterized by a build-up of free electrons in the vacuum of an accelerator. These electrons have several sources and can be amplified by the beam's electromagnetic field. The electrons can then have various deleterious effects on the beam, including vacuum activity, detector backgrounds, beam instability, and heat deposition.

Detailed study of the electron cloud is only a recent interest in research. It has the curious behavior where it is not an issue at all for low-intensity beams, but can be quite severe at high-intensity. Additionally, the electron cloud has no obvious signature in conventional beam instrumentation.

Our interest is in the design and operation of new, high-current machines. Particularly, at Fermilab we are proposing a new, high-power linear accelerator that will lead to multi-megawatt proton beams at a few GeV, and from an upgraded Main Injector at 120 GeV. The electron cloud is at the top of our mind as an issue for higher beam current in the Main Injector. Also, the electron cloud is an issue for other accelerators such the LHC and the proposed ILC. We note that the electron cloud will be an issue for all future high-current positive beam accelerators, as well as electron machines where there is substantial synchrotron radiation.

## THE PROJECT X UPGRADE

One component of the proposed Project X is to increase the beam current of the Main Injector by three to four times. This will be accomplished with a new multi-mission linac and/or synchrotron.

The Main Injector presently provides high-power proton beams for antiproton and neutrino production. The present power is now approximately 400 KW with beam intensities of 4-5e13 protons (6-10e10 protons per 53 MHz bunch). In the near future the accelerator will be upgraded to provide 700 kW for the NOvA neutrino experiment, but this will be accomplished with only a nominal increase in beam intensity (the cycle time will be reduced).

The Project X upgrade, however, will increase the bunch intensity to about 30e10 protons per bunch. This increase will require significant upgrade of the RF accelerating cavities and may aggravate other existing instabilities. For the electron cloud, there is the potential for the situation to be much worse: a threshold may be crossed where the electron cloud increases many-fold. We are trying to head off this problem by extensive study of the present beams in the Main Injector.

## MODEL OF THE ELECTRON CLOUD FOR FERMILAB

The precise formation of the electron cloud can vary between accelerators as the seeding and amplification process can vary substantially. Here, we outline the process for the Main Injector.

The Main Injector beam is a train of 500 bunches separated by 19 ns each. The bunch length varies during the acceleration cycle between about 1 and 8 ns. The transverse size also varies with a sigma of 1 - 5 mm. The maximum operating bunch intensity is approximately 10e10 protons.

As the bunches start to pass a particular location they can start to produce electrons through two processes:

- Residual gas ionization, where the residual gas in the vacuum is ionized by the beam. This process produces (order of magnitude) 1 electron per meter, per torr, per proton, per pass). The Main Injector vacuum is typically 1e-9 torr, so this will produce about 100 electrons per meter, per bunch crossing (a rather small number).
- Proton losses, where beam is lost on the beam pipe and looses electrons as it travels through the material. If impacting at grazing angles this process can produce 100s of electrons per proton. However, the losses in the Main Injector occur predominantly at the start of the accelerating cycle and are localized to a few areas of the accelerator. This process will thus contribute negligibly compared to residual gas ionization.

Another potential source is synchrotron radiation. It is negligible here, but is significant for electron and positron machines. The LHC is of high enough energy that there is substantial seeding from the synchrotron radiation – high-energy muons accelerators would also be at risk.

In certain machines the seed electrons could be a problem themselves, but they are negligible for the Main Injector. Instead, they are only a problem when amplified. The amplification process is mediated by two steps:
- Heating of the electrons in the electrostatic potential of the beam, which appears non-adiabatically.
- Multiplication through secondary emission on the beam pipe wall.

In detail, the electrons are accelerated towards the center of the beam pipe where the beam is. Next, the beam bunch passes before the electrons decelerate on the other side of the beam, so they gain net energy. They continue along ballistic paths while between bunches and impact with the beam pipe wall. Depending on the chemistry of the wall surface and the energy of the electrons, a number of secondary electrons are produced that may be greater than the incident number (the secondary electron yield – SEY). Those electrons are then re-accelerated by the subsequent bunch and the process continues. If the number of electrons is greater, the net number grows exponentially with bunch passings until the electric field of the electrons screen the beam's field. As this point the number of electrons is the same order as that of the protons.

If however, the secondary emission can be reduced somehow, the exponential growth can be suppressed and the electron cloud mitigated. It is known that the Main Injector beam pipe material (SS 316L) initially produces two electrons for each incident 400 eV electrons (fewer for other energies). This number can be reduced through coatings or beam scrubbing. The produced electrons typically have energies of 1-10 eV.

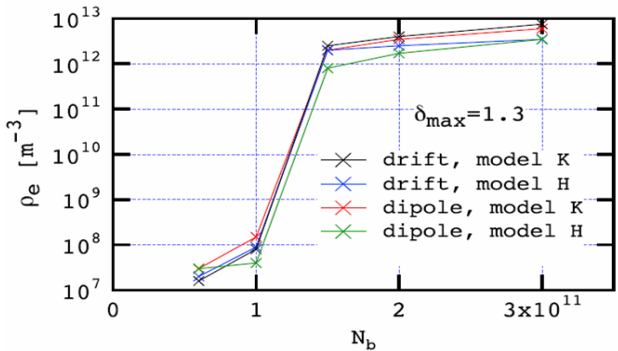

Figure 1: POSINST simulation of electron cloud equilibrium density versus charge per bunch.

Quantitative prediction of electron cloud formation requires the use of computer simulation of the acceleration and amplification processes. We have used several codes, the first of which was POSINST [1]. Using a "typical" value of the maximum secondary emission of 1.3, that simulation predicted a 100,000 increase of electron density when increasing the bunch intensity from 10e10 to 30e10, as the Project X upgrade would. This precipitous threshold would occur just above 10e10 protons and is very concerning for any future operation.

*Critical Model of Electron Cloud Formation*

Here, we develop a simple to explore why such a large threshold might be expected.

Consider a critical parameter $\kappa$. This parameter is the ratio of electrons at a particular time in a bunch passing, to the same time of the previous passing (excluding primary production). $\kappa$ then encodes all the effects of accelerations, secondary production, and the slower loss of electrons before reheating by the next bunch.

$$N_b = \kappa \times N_b + P$$

Where $N_b$ is the number of protons in bunch $b$ and $P$ is the number of primary electrons produced. There is no straightforward equation for $\kappa$, but we can expect it to grow somewhat smoothly with beam intensity. We can separate this into two regimes, whether $\kappa$ is greater or less than 1.

If $\kappa < 1$, there is no exponential growth, but there is still some accentuation of the primary electrons. This is because the secondary production process essentially lengthens the lifetime of the electrons, even though it does not increase them. In this case an equilibrium number of electrons will be reached after a large number of bunch passes:

$$N_{eq} = \frac{P}{1-\kappa}$$

So if $\kappa$ is close to one the equilibrium density could be significantly larger than that produced through primary production itself. In machines where $P$ is a significant fraction of the beam (such as electron or positron synchrotrons) this can lead to the build-up of a damaging cloud even if there is no runaway amplification. For the Main Injector, however, $P$ is extremely weak so the $N_{eq}$ is negligible for all values of $\kappa < 1$.

On the other hand, if $\kappa > 1$ there is no equilibrium value in our simple model. Instead there is exponential growth which is only quenched by the space charge of the electrons themselves. That quenching will produce another equilibrium value based on $\kappa$. Simulation is required for quantitative evaluation, but we expect it to be on the same order as, but smaller than the beam density.

Therefore, in proton machines where the primary forcing is weak, the transition can be very strong from relatively no electrons to a very intense cloud.

## RESEARCH APPROACH

Our program of research for the electron cloud is focused on developing a strategy for the Project X upgrades by performing experiments in the Main Injector at present intensities and using simulation to be confident in our extrapolation to higher intensity. Our default approach with Project X is to coat most or all of the Main Injector with a low secondary yield surface (probably TiN, or a carbon-based surface). Our studies aim to validate the mitigative effect of the various coatings, and to study the electron cloud dynamics so that we can be confident that effect is maintained at higher intensities. Also, the coating will be time-intensive and somewhat expensive to apply, so we want to know how much and how well the ring needs to be coated.

As will be shown, we have observed the beginning of a threshold in the Main Injector and monitored the evolution of the secondary yield of the beam pipe. We have developed new instrumentation and have worked at benchmarking our experience to that of other labs (predominantly Cornell).

## INITIAL ELECTRON CLOUD EVIDENCE

A first search for evidence of the electron cloud was started in 2006 using existing instrumentation. A large number of vacuum bursts were found around the ring. These bursts were transient in time, but not large enough to activate the machine protection systems. The pressure increases were consistent with electron cloud activity, and seemed to show a clear threshold, but better evidence was needed. A RFA (retarding-field analyzer) was borrowed from Argonne and installed in a drift section of the Main Injector in 2007.

### First Observation of Formation Threshold

The RFA showed immediate results in the high-intensity studies then underway. At that time, the typical bunch intensity ion the Main Injector was only 6e10 protons, but occasional studies would bring that as high as 10e10. During those studies the cloud could be observed with the RFA, but not at low-intensity.

While the beam intensity was being pushed up, all beam cycles were measured over the course of a week to take advantage of the variation of intensity. This allowed mapping out a clear threshold at 26e12 total protons. The electron current was zero below this level, and increase linearly above it.

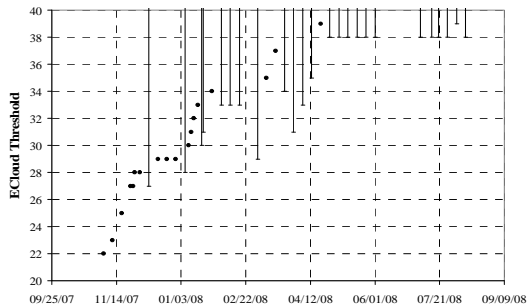

Figure 2: Evolution of the initial cloud threshold.

The beam threshold was found to vary with time. Particularly, the threshold increased whenever a new, higher intensity became part of regular operations. This led us to conclude that it was the scrubbing of the electron cloud itself that reduced the secondary electron emission yield of the beam pipe surface, increasing the formation threshold.

### Study of Beam Instability Threshold

After observing the formation of the cloud, we needed to understand if it was inducing any instability in the beam. This task is made difficult because the Main Injector beam at marginal intensities already suffers a coupled-bunch instability from the resistive wall effect. A digital bunch-by-bunch damping system is necessary to accelerate beam of any substantial intensity. This damping system would also suppress any coupled-bunch oscillation induced by the electron cloud.

To test whether the electron cloud was contributing substantially to the instability, a study of the instability threshold was performed. In this case, beam of certain intensity was accelerated for many cycles. The damper gain was reduced until the beam went unstable. The gain needed to counteract the resistive wall should vary linearly with beam intensity. If the electron cloud contributed substantially, it would manifest as a variation from this linear relation at the higher intensity where the cloud was observed. Instead, we found only the predictably linear relation, so we concluded that any instability induced by the electron cloud was still much smaller than the existing resistive wall.

## THE 2009 EXPERIMENTAL STATION

A new experimental station was installed in the Main Injector during the 2009 accelerator shutdown. This dedicated area allowed several types of study:
- Validation of low SEY coatings.
- Measurement of formation threshold.
- Measurement of conditioning.
- Study of dynamics throughout acceleration cycle.
- Energy spectrum measurement.
- Instrumentation tests and comparison.
- Study of longitudinal dynamics.

The chief components of the station were two identical 6"Ø, 1 m length cylindrical beam pipe. These pipes, installed in series, can have different coatings applied and be tested side-by-side with identical instrumentation and beam conditions. Each pipe had 3 RFA ports and was installed between microwave antennas and absorbers.

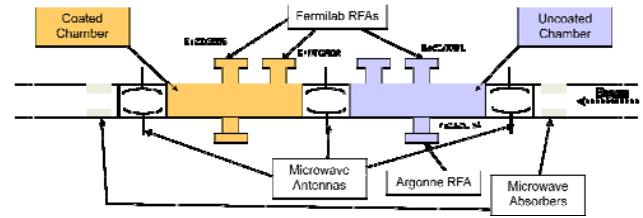

Figure 3: Layout of the 2009 experimental upgrade.

### TiN Coating

The first coating to be tested was TiN vs. the raw Stainless 316L. Several procedures exist to apply TiN.

We sent our pipe to Brookhaven to be coated there in the process developed for RHIC.

*Electron Detectors (RFAs)*

We developed a new RFA (retarding field analyzer) style to be installed in addition to the borrowed Argonne RFA. Our RFA had several improvements that were useful for the Main Injector conditions:
- Removal of the extra ground grid, increasing collection efficiency by about 9%.
- Maximizing the collection area
- Designing electrodes to be more hermetic, avoiding the loss of electrons through the sides, and to better operate as an energy filter.
- Introduction of amplification in the tunnel with moderately radiation-hard components.

This detector has sensitivity above the noise floor a factor of several hundred times better than the previous installation, allowing close inspection of the start of the threshold.

## MEASUREMENTS WITH THE UPGRADED STATION

Data from the several RFAs are collected into the Fermilab accelerator controls system (ACNET). This data can be viewed online, within a cycle with a ~ 3 kHz bandwidth. The peak current during each accelerator cycle is stored to disk along with useful correlative data, such as beam intensity, beam mode, bunch length.

Upon start-up, we observed very definite signals at rather low intensity. This is thought to be because the beam pipe starts with an SEY of ~ 2.0, which requires exceedingly little electron heating to produce the runaway cloud formation. However, the surface chemistry is rapidly affected by the intense electron bombardment to a SEY of much lower value. This conditioning is key to the operation of the accelerator and has been observed to evolve for months. The TiN current was initially lower than the SS, but still significant. Both decreased rapidly with conditioning.

We note that the electron bombardment also causes significant gas desorption, causing occasional vacuum bursts when the beam pipe surface is still young. It is not clear whether the electron cloud is the dominant process in vacuum processing of the Main Injector and other similar accelerators, but other previously posited mechanisms (such as heating from image currents) seem much weaker.

*Threshold Measurement*

Our chief benchmark for the electron cloud severity is the beam intensity threshold at which the electron cloud starts to form. To measure this, we collect all the data from a period of time (day or week). For each cycle, the beam intensity and electron flux are compared. All of these points tend to lie on a smooth curve that is zero for the majority of the intensity range, and then increases rapidly at a certain value of beam intensity. The curve is fit to a simple formula and a benchmark beam intensity ($x_0$) is extracted at which the cloud flux increases over a certain value. We call this benchmark a threshold, though the transition is smooth. This value is different than the threshold observed in the previous installation where the smooth turn-on could not be observed.

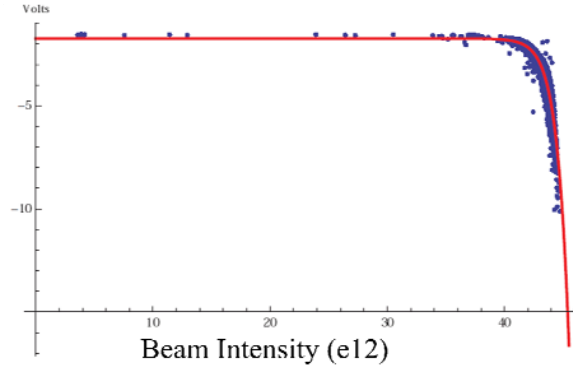

Figure 4: A day's worth of data comparing the detector signals (volts) to beam intensity. Each acceleration cycle is a point and the fit line is shown.

*Evolution of Thresholds*

Most important to use is how the threshold evolves with conditioning. We expect that the conditioning is not just a factor of time, but is more correlated with the total electron fluence or dose on the surface. We measure the total dose by integrating the flux measured on the detectors. The conditioning history of the TiN and stainless pipes are shown in Fig. 5. The threshold is plotted as a function of the integrated dose to that date.

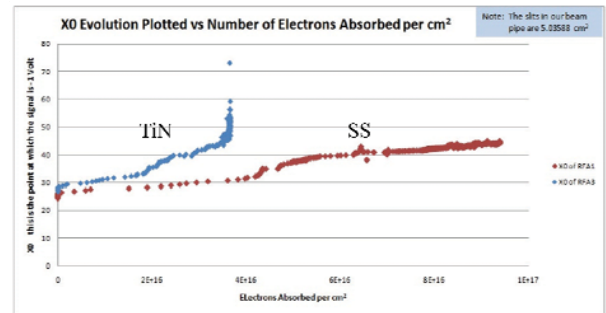

Figure 5: Conditioning history of TiN versus Stainless. The TiN conditions much more quickly than the stainless and to a higher value.

Not only does the TiN condition much more quickly and to a higher value than the stainless, but after a few months almost no electron cloud was observed; while for stainless the cloud was always visible at high intensity. This shows a typical behavior where the pipe will condition to the maximum beam intensity available. Unfortunately, we cannot be certain of the behavior at higher intensities without achieving them, or studying in more detail the SEY of the material.

*Note on Rate of Conditioning*

The performance of TiN (and even stainless) in the Main Injector has been superior to that at other proton synchrotrons, such as the SNS and PSR. We attribute this difference to a much higher electron dose at the Main Injector. To first order, the electron flux on the wall is proportional to the cloud density and the *total* number of bunches passing the location (each bunch causes nearly all of the electrons to collide with the wall once).

For the Main Injector, the total number of bunches is vastly larger. It operates a 53 MHz system, while the PSR and SNS operate at ~ 1 MHz. Additionally, the Main Injector accelerates its beam, so high-intensity beam persists for about 1 s of a 2 s cycle (50% duty factor). The SNS and PSR only accumulate beam and then immediately extract, so the duty factor for high-intensity beam is much lower (~ 0.1%). As such, the Main Injector will see tens of thousands more electron dose for the same cloud density.

*Amorphous Carbon Coating*

In 2010 we installed a new test chamber with an amorphous carbon coating provided by CERN. CERN is pursuing the carbon coating because their tests have indicated it requires less initial conditioning than the TiN.

Our experiment with the carbon is still ongoing. So far, we can say that it showed similar behavior to TiN for the first few weeks (that is, it required conditioning).

A vacuum leak occurred at that few weeks into the run near the carbon station. It caused significant worsening of the cloud density at the nearest RFA. Its behavior was worse than the steel for several months after, until it finally recovered. Thus, in our experience the carbon appears to be very susceptible to air and/or vacuum issues.

## MICROWAVE MEASUREMENTS

Microwaves can be used to measure the electron cloud density through the phase modulation it causes on a carrier wave. This technique has been popular recently as it can allow a distributed measurement with very little instrumentation (typically BPMs can be used as antennas). We implemented out own procedure which allowed time-resolving of the cloud behavior by directly detecting the phase shifts [2], instead of relying on sidebands as other measurements had.

*Issues with Previous Measurements*

Unfortunately, the normalization of the phase modulation to produce a cloud density was always problematic when compared to theory. Additionally, the behavior was erratic in some locations.

We concluded that there were several flaws with the method, mostly because of reflections caused throughout the accelerator. Theses reflections could be caused by small deviations in the surface as the microwave is typically just above cutoff. The reflections produce many separate paths between the transmitter and receiver. We found a 20 dB reduction of transmission when ferrite absorbers were put around the area of interest. This indicates that almost all the phase shift measured was from these extra paths. Those paths were much larger and passed through an unknown set of elements. Therefore the measurement could not be localized to an area, nor could any normalization be applied. We believe this problem affects all other microwave transmission measurements of the electron cloud.

*Concept for an Improved Procedure*

We are designing a new microwave test cell where the wave will be confined by reflectors around the area of interest. This will allow the carrier wave to be further above cutoff, but still gain signal by the controlled multiple reflections within the cavity.

## SEY TEST STAND

We plan to install an in-situ measuring device for the SEY. This device was developed at SLAC, and then with Cornell. This will allow us to directly measure the SEY of material samples after irradiation with the beam. This SEY value will be used to further constrain simulations and aid our extrapolation to Project X intensities.

## SUMMARY

We have studied the electron cloud extensively at the Fermilab Main Injector. We find the appearance of the cloud over a threshold, but so far its effects on the beam are benign. We have found that TiN and amorphous-carbon coated chambers show superior performance to stainless steel, though the carbon may be somewhat fragile. We have developed new instrumentation, including an RFA and time-resolved microwave phase measurements. We are working to resolve the issues with the microwave measurement and to directly measure the SEY of irradiated materials. When these experiments are complete and combined with simulation, we will be able to make definite predications for Project X and specify the mitigations needed.

## ACKNOWLEDGEMENTS

This work was the product of many people's work at Fermilab and elsewhere. At Fermilab, substantial contributions come from M. Backfish, D. Capista, J. Crisp, K. Duel, N. Eddy, I. Kourbanis, C. Y. Tan, C. Thangaraj, L. Valerio, and X. Zhang. Also, M. Furman from LBNL, C. Harkay from ANL, M. Pivi from SLAC, M. Palmer *et al.* from Cornell.